\documentclass[11pt,a4paper,oneside,onecolumn]{article}
\usepackage[utf8]{inputenc}
\usepackage[width=17.00cm, height=25.00cm]{geometry}

\usepackage{latexsym}
\usepackage{graphicx}
\usepackage{mathptmx}
\usepackage[T1]{fontenc}

\usepackage{amsmath}
\usepackage{amsfonts}
\usepackage{amssymb}
\usepackage{amsbsy}
\usepackage{amsthm}

\usepackage{multirow}
\usepackage[nolist]{acronym}

\usepackage[pdftex,colorlinks=true,urlcolor=blue,citecolor=black,anchorcolor=black,linkcolor=black]{hyperref}

\def\email#1{\href{mailto://#1}{#1}}
\makeatletter
\let\@internalcite\cite
\def\cite{\def\@citeseppen{-1000}%
    \def\@cite##1##2{(##1\if@tempswa , ##2\fi)}%
    \def\citeauthoryear##1##2##3{##1 ##3}\@internalcite}
\def\citeNP{\def\@citeseppen{-1000}%
    \def\@cite##1##2{##1\if@tempswa , ##2\fi}%
    \def\citeauthoryear##1##2##3{##1 ##3}\@internalcite}
\def\citeN{\def\@citeseppen{-1000}%
    \def\@cite##1##2{##1\if@tempswa, ##2)\else{}\fi}%
    \def\citeauthoryear##1##2##3{##1 (##3)}\@citedata}
\def\citeA{\def\@citeseppen{-1000}%
    \def\@cite##1##2{(##1\if@tempswa , ##2\fi)}%
    \def\citeauthoryear##1##2##3{##1}\@internalcite}
\def\citeANP{\def\@citeseppen{-1000}%
    \def\@cite##1##2{##1\if@tempswa , ##2\fi}%
    \def\citeauthoryear##1##2##3{##1}\@internalcite}
\def\shortcite{\def\@citeseppen{-1000}%
    \def\@cite##1##2{(##1\if@tempswa , ##2\fi)}%
    \def\citeauthoryear##1##2##3{##2 ##3}\@internalcite}
\def\shortciteNP{\def\@citeseppen{-1000}%
    \def\@cite##1##2{##1\if@tempswa , ##2\fi}%
    \def\citeauthoryear##1##2##3{##2 ##3}\@internalcite}
\def\shortciteN{\def\@citeseppen{-1000}%
    \def\@cite##1##2{##1\if@tempswa, ##2\else{}\fi}%
    \def\citeauthoryear##1##2##3{##2 (##3)}\@citedata}
\def\shortciteA{\def\@citeseppen{-1000}%
    \def\@cite##1##2{(##1\if@tempswa , ##2\fi)}%
    \def\citeauthoryear##1##2##3{##2}\@internalcite}
\def\shortciteANP{\def\@citeseppen{-1000}%
    \def\@cite##1##2{##1\if@tempswa , ##2\fi}%
    \def\citeauthoryear##1##2##3{##2}\@internalcite}
\def\citeyear{\def\@citeseppen{-1000}%
    \def\@cite##1##2{(##1\if@tempswa , ##2\fi)}%
    \def\citeauthoryear##1##2##3{##3}\@citedata}
\def\citeyearNP{\def\@citeseppen{-1000}%
    \def\@cite##1##2{##1\if@tempswa , ##2\fi}%
    \def\citeauthoryear##1##2##3{##3}\@citedata}
\def\@citedata{%
    \@ifnextchar [{\@tempswatrue\@citedatax}%
                  {\@tempswafalse\@citedatax[]}%
}

\def\@citedatax[#1]#2{%
\if@filesw\immediate\write\@auxout{\string\citation{#2}}\fi%
  \def\@citea{}\@cite{\@for\@citeb:=#2\do%
    {\@citea\def\@citea{, }\@ifundefined
       {b@\@citeb}{{\bf ?}%
       \@warning{Citation `\@citeb' on page \thepage \space undefined}}%
{\csname b@\@citeb\endcsname}}}{#1}}%

\def\@citex[#1]#2{%
\if@filesw\immediate\write\@auxout{\string\citation{#2}}\fi%
  \def\@citea{}\@cite{\@for\@citeb:=#2\do%
    {\@citea\def\@citea{; }\@ifundefined
       {b@\@citeb}{{\bf ?}%
       \@warning{Citation `\@citeb' on page \thepage \space undefined}}%
{\csname b@\@citeb\endcsname}}}{#1}}%

\def\@biblabel#1{}
\makeatother

\newdimen\bibindent
\usepackage{fancyhdr}
\fancypagestyle{plain}{
	\fancyhead[L]{\textit{~\\Accepted version of the paper presented at the 2024 Winter Simulation Conference\\DOI: \href{https://doi.org/10.1109/WSC63780.2024.10838852}{10.1109/WSC63780.2024.10838852}\\979-8-3315-3420-2/24/\$31.00 \copyright~IEEE~2024}}
}

\newtheoremstyle{wsc}
{3pt}
{3pt}
{}
{}
{\bf}
{}
{.5em}
{}

\theoremstyle{wsc}

\begin{acronym}
	\acro{BESS}{battery energy storage system}
	\acro{EV}{electric vehicle}
	\acro{LoD}{Level of Detail}
	\acro{MiD}{Mobility in Germany}
	\acro{PV}{photovoltaic}
	\acro{SCR}{self-consumption rate}
	\acro{SOC}{state of charge}
	\acro{SSR}{self-sufficiency rate}
\end{acronym}

\newcommand{\ignore}[1]{}

\begin{document}

%
%

\title{Analyzing the Impact of Electric Vehicles on Local Energy Systems using Digital Twins}

\author{
	Daniel René Bayer and Marco Pruckner\\
	{\small  Modeling and Simulation, University of W\"urzburg, 97074 W\"urzburg, GERMANY}
}

\date{}

\maketitle

\vspace{-12pt}

%
%
%
%

\section*{ABSTRACT}
The electrification of the transportation and heating sector, the so-called sector coupling, is one of the core elements to achieve independence from fossil fuels.
As it highly affects the electricity demand, especially on the local level, the integrated modeling and simulation of all sectors is a promising approach for analyzing design decisions or complex control strategies.
This paper analyzes the increase in electricity demand resulting from sector coupling, mainly due to integrating electric vehicles into urban energy systems.
Therefore, we utilize a digital twin of an existing local energy system and extend it with a mobility simulation model to evaluate the impact of electric vehicles on the distribution grid level.
Our findings indicate a significant rise in annual electricity consumption attributed to electric vehicles, with home charging alone resulting in a 78\% increase. However, we demonstrate that integrating photovoltaic and battery energy storage systems can effectively mitigate this rise.

\section{INTRODUCTION}
\label{sec:intro}
	
	In recent years, there have been significant advancements in energy systems research, notably in two key areas: The utilization of digital twins based on smart meter data of energy systems and an increased focus on analyzing the impacts of sector coupling on the (electrical) energy system.
	Sector coupling, i.e., the integration of various sectors like heating and mobility into the energy system, holds global significance for the pursuit of decarbonization \shortcite{2018_Brown_SectorCouplingAndTransmissionEurope}.
	Furthermore, beyond decarbonization objectives, sector coupling presents opportunities to diminish reliance on imported fossil fuels, potentially averting energy crises like the European gas crisis of 2022 \shortcite{2023_Mannhardt_MitigationStrategiesEuropeanGasCrisis}.
	In the context of sector coupling, this paper focuses on integrating \acp{EV} into the energy system.
	The integration of \acp{EV} is critical at a local level, especially in cities, as uncontrolled charging behavior can lead to a substantial increase in the peak demand of the energy system on the local as well as national level \shortcite{2022_Strobel_JointAnalysis_EV_Impact}.
	
	In the recent past, digital twins have emerged as a powerful technology for addressing queries concerning the future of urban living, including local energy systems.
	Even though the concept of digital twins initially emerged in product lifecycle management, the urban space is currently the most relevant domain for applying digital twins, according to \shortciteN{2022_Botin_DigitalTwinChallengesReview}.
	While there exist varied interpretations of the prerequisites for digital twins of energy systems, a common thread is the imperative for these models to accurately mirror real-world systems based on data extracted from diverse sources \shortcite{2021_Bhatti_EV_DTs}.
	In the context of energy systems, this could be the combination of smart meter data reflecting the actual electricity consumption with a high temporal resolution and building-related data like roof areas, enabling precise modeling of possible future \ac{PV} installations.
	
	However, integrating mobility aspects into an energy system's digital twin, particularly \ac{EV} driving and charging behaviors, poses a unique challenge, as mobility data is required at the level of the individual buildings in order to comply with the digital twin paradigm.
	Whereas obtaining smart meter data and building-related data to build a digital twin of the entire energy system only requires the existence of smart meters, including the communication infrastructure and a cartography service that many government agencies nowadays offer, obtaining the mobility data of \textit{all} city inhabitants is almost impossible as the actual movement profiles of all vehicles would be required.
	Although modern vehicles already record these data, they are is not publicly available.
	Therefore, it is infeasible to build a digital twin that reflects the existing local energy system and the actual mobility needs of citizens.
	To address this gap, we present a novel hybrid approach combining a mobility demand simulator \cite{2023_Strobel_OMOD} and a digital twin of a city’s energy system \cite{2023_Bayer_DT_LocalEnergySystem}.
	We aim to illuminate the crucial role of digital twins in assessing the complexities of future energy systems.
	In this paper, we answer the following research questions:
	How to process the output of a people-centered mobility demand generator to get spatial-disaggregated \ac{EV} driving profiles and how to integrate these \ac{EV} profiles in a digital twin of a city's energy system?
	To demonstrate the power of the combination of a real energy system's digital twin with a mobility demand simulation, we ask which share of \ac{EV} charging demand can be covered by a residential rooftop \ac{PV} installation in a realistic scenario while considering the unique characteristics of each household assuming that the \acp{EV} are only charged at home.

	The rest of the paper is organized as follows: The related work is presented in Section~\ref{sec:related_work}.
	In Section~\ref{sec:methodology}, we present the methodology, i.e., combining a digital twin of an energy system with a mobility simulation on the level of individual citizens.
	In Section~\ref{sec:results}, the results of the mobility simulation and the possibility of using rooftop \ac{PV} installations to cover the \ac{EV} charging demands are analyzed at the level of individual buildings.
	Finally, we discuss the results and the limitations of the methodology in Section~\ref{sec:discussion_and_limitations} and conclude the paper in Section~\ref{sec:conclusion}.

\section{RELATED WORK}
\label{sec:related_work}

In recent years, integrating digital twin technologies has emerged as a promising approach for enhancing the synergy between mobility and electricity systems.
\citeN{2021_Papyshev_CityDigitalTwins_UrbanMobility} address the challenge of generating synthetic mobility data within a digital twin framework to mitigate privacy concerns associated with real data usage.
\shortciteN{2022_Wang_MobilityDigitalTwinReview} present a digital twin for representing mobility systems, encompassing both vehicles and human beings within a cloud-based environment.
Their focus lies on real-time applicability. The construction of an urban digital twin, including mobility data, is also presented by \shortciteN{2021_Lee_UrbanMobilityDigitalTwin}.
They also integrate dynamic mobility data, but in contrast to the previous paper, focusing on vehicle and pedestrian detection.
\shortciteN{2021_White_SmartCityDigitalTwin} conceptualize urban digital twins as multi-layered constructs, incorporating diverse urban information to facilitate holistic simulations and optimizations.
Their proposed use cases, including mobility demand optimization and long-term planning, highlight the versatility and utility of digital twin technologies across various urban domains.
A conceptual novelty of \shortciteN{2021_White_SmartCityDigitalTwin} is the understanding of a digital twin as the uppermost layer of a multi-layered virtual representation of the urban area ranging from the level of the buildings, over the infrastructure and mobility up to the digital layer on top.
Moreover, digital twins can also be utilized to visualize and analyze urban traffic dynamics, offering valuable insights into their implications for air quality, as demonstrated by \citeN{2022_Bachechi_UrbanMobilityDTs}.
In the transportation sector, digital twins hold promise for optimizing fuel consumption and enhancing transportation efficiency, as emphasized by \shortciteN{2023_Kajba_DTsInTransportationAndEnergyReview}.
Their research underscores the potential of digital twin applications in addressing key challenges facing modern transportation systems, such as energy consumption and environmental sustainability.

Based on the results of \shortciteN{2013_Munkhammer_EVHomeChargingWithPV}, we conclude that the analysis of fulfilling the \ac{EV} charging demand by a residential \ac{PV} installation is a vital topic to analyze, especially on a building-resolved level, as one of their findings is that the \ac{EV} charging demand may not necessarily temporarily coincide with the \ac{PV} generation.
For instance, \shortciteN{2022_Martin_CoverEVDemandWithPV} analyze the ratio of the \ac{EV} electricity demand that can be covered by an individual rooftop \ac{PV} installation in Switzerland, reporting that only 15\% of the \ac{EV} electricity consumption can be covered with \ac{PV} generation using immediate (i.e., uncontrolled) charging.
Changing the charging strategy is one way to increase the usable \ac{PV} generation for \ac{EV} charging.
For instance, \citeN{2024_Benz_SmartEVChargingVsImmediateCharging} compare such an immediate \ac{EV} charging strategy with more sophisticated smart charging strategies based on linear optimization for a corporate parking garage.
When only focusing on immediate charging, increasing the \ac{PV} generation used for \ac{EV} charging seems easier in parking garages than for home charging. 
For instance, \shortciteN{2023_Sing_EVChargingWithRenewableAndStorage} present that even without smart charging over 60\% of \ac{PV} generation can be used for direct \ac{EV} charging for a parking garage in California.
This is a notably higher proportion compared to home charging \shortcite{2022_Martin_CoverEVDemandWithPV}.
Meanwhile, a seasonal analysis by \shortciteN{2021_Cieslik_EV_and_PV_Combination} presents that \ac{PV} generation is sufficient to cover most of the \ac{EV} charging demand at home between April and August in an exemplary European country.
Despite the potential benefits of utilizing rooftop \ac{PV} installations for \ac{EV} charging, concerns regarding the peak load increase persist.
In this context, \shortciteN{2022_Strobel_JointAnalysis_EV_Impact} note that the question of overloads highly depends on the used control strategy for \ac{EV} charging.
Finally, \shortciteN{2013_Paevere_SpatioTemporalModelingEVCharging} consider the \ac{EV} charging demand and its impacts on the peak loads of the buildings for a town in Australia using a fine-grained, spatially resolved model presenting notable impacts on the peak load on the building level.

\subsection{Research Gap}
Existing literature predominantly focuses on the development of digital twins with a focus on real-time applications such as those demonstrated by \shortciteN{2022_Wang_MobilityDigitalTwinReview} or \shortciteN{2021_Lee_UrbanMobilityDigitalTwin}, or they aim to visualize mobility patterns \cite{2022_Bachechi_UrbanMobilityDTs}.
While these papers address questions pertinent to short-term scenarios, \shortciteN{2021_White_SmartCityDigitalTwin} focus on using urban digital twins, including a representation of mobility for long-term planning.
Nevertheless, all cited papers require mobility data, which is hard to obtain on the level of all city residents.
Therefore, combining an energy system's digital twin with a mobility simulation on the individual level seems promising and fast to implement.
Thus, we first present a methodology for combining these two concepts, which is required for various use cases.
Thereupon, we evaluate such an use case, which is evaluating the possibility of covering \ac{EV} home-charging demand by a rooftop \ac{PV} installation over all existing buildings in a city incorporating the building-related circumstances.
These include the roof areas and shapes limiting the addable \ac{PV} installation size or the existing electricity consumption of the residential buildings as it reduces the available \ac{PV} generation for \ac{EV} charging.
This is important as existing analyses that do not rely on digital twins, like \shortciteN{2022_Martin_CoverEVDemandWithPV}, do not consider the existing electricity consumption of the buildings where  \acp{EV} and \ac{PV} installations are added.

\section{METHODOLOGY}
\label{sec:methodology}

\begin{figure}[htb]
		{
			\centering
			\includegraphics[width=0.85\textwidth]{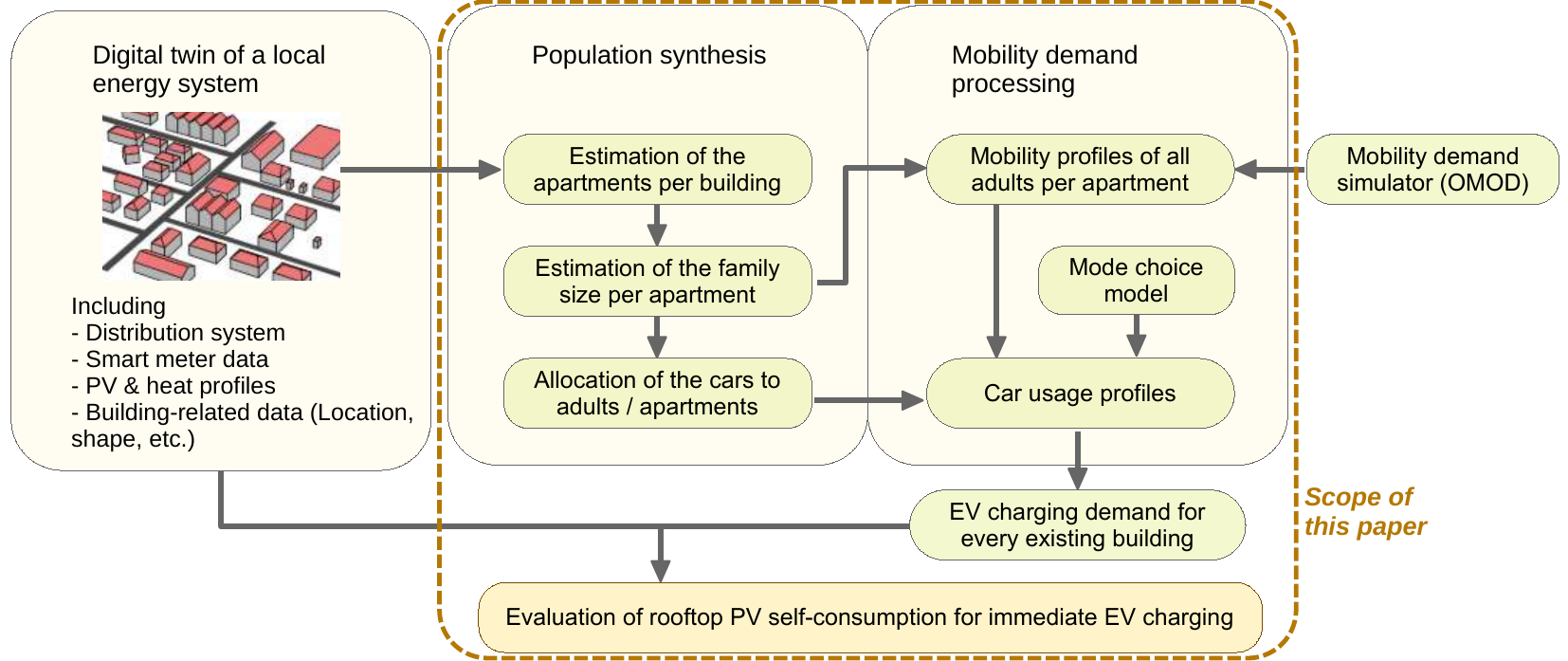}
			\caption{Methodology and scope of this paper.\label{fig:methodology}}
		}
	\end{figure}

	In this section, we first describe the methodology of sampling the population of a given city based on building-related parameters.
	Thereupon, we describe the processing of the output of a mobility demand simulation on the individual person level and how to aggregate them to vehicle usage profiles, incorporating a mode choice model.
	In a second step, we present how we extend an existing implementation of a digital twin to include the driving profiles, assuming the sampled vehicles would be \acp{EV}.
	A general overview of the presented methodology and the general scope of this paper is illustrated in \autoref{fig:methodology}.
	
\subsection{Structure of the Digital Twin}
	This paper utilizes a digital twin of an energy system of a town in Southern Germany that is described in detail in \citeN{2023_Bayer_DT_LocalEnergySystem}.
	This digital twin accurately replicates the current energy system down to individual buildings, encompassing substations, the individual buildings including their electricity consumption based on smart meter data and additional building-related information like their volumes, locations or shapes.
	Its primary focus is the integration of diverse data sources to realistically model future scenarios, particularly those involving increased sector coupling.
	These scenarios include increasing rooftop \ac{PV} penetration, additional \acp{BESS}, and the adoption of heat pumps.

\subsection{Used Dataset}
	We obtained the smart meter time series on the building level and the topology of the distribution system from the local utility company of Ha{\ss}furt, Germany.
	The dataset contains the smart meter data with an hourly resolution from multiple years, including 2021, which will be the basis of the later analysis.
	They are combined with cadastral data including roof shapes and orientations for the simulation of new \ac{PV} installations.
	Moreover, we have additional information about the existing \ac{PV} installations and \ac{BESS}, including \ac{PV} generation profiles with an hourly resolution.
	
	In order to combine the existing building and energy data with information on mobility behavior, the paper includes additional survey data.
	Firstly, we conducted a survey in which we invited all households in Haßfurt to participate.
	Details on the questions and the results can be found in Section \ref{sec:methodology_survey_charging_choice}.
	Secondly, we use the results of the \acf{MiD} dataset, which is a travel survey with data collected in 2017 \cite{2017_MiD}. 
	Finally, over 150{,}000 households participated. 
	The \ac{MiD} basically asks the residents of all participating households to report all their trips on a randomly chosen day.
	For each trip, the means of transport used, the departure and arrival time, as well as the reason for the trip are recorded.

\subsubsection{Survey Details}
\label{sec:methodology_survey_charging_choice}
	We conducted a survey among all residential customers/households of the utility company with a response rate of 17\% ($n = 1101$).
	For the validation of the survey, we compare two different building-related parameters of the buildings from which we have survey results with all residential buildings in the considered town.
	Regarding the building volume as the first tested parameter, the null hypothesis of the t-test of different means cannot be rejected $(t(5824) = -0.36, p = 0.72, d = -0.01)$.
	For the second comparison, we use the building type (detached or semi-detached building, townhouse or other) of all residential buildings in the considered area as given in the latest available census with the answers as given in the survey.
	The null hypothesis of the chi-squared test on equal relative frequencies of the building types cannot be rejected $(\chi^2(3,1101) = 0.023, p > 0.99)$.
	
	In the survey, we asked, among other things, about the number of apartments in the building, the number of adults and children, and the number of vehicles in the questioned household.
	For every vehicle per household, we asked for the average number of days per week when the vehicle is used and the estimated daily vehicle driving distance (only for the days with vehicle usage).
	Moreover, we included questions on the charging behavior of \ac{EV} users.
	Therefore, we asked how often the \ac{EV} is connected to the charging station when the driver arrives at home.
	The results indicate that almost 50\% of the \acp{EV} are only connected every fourth time the \ac{EV} arrives at home.

\subsection{Population Synthesis}
	Since the existing digital twin does not include information about how many flats are present per building, or how many people live there, or how many vehicles they have, we have to estimate this information based on the available data and our conducted survey.
	Therefore, we use a two-stage approach:
	First, we train a decision tree to predict the building type (i.e., single-family building, two-family building or apartment tower) based on the survey results as ground truth. 
	The input parameters for the decision tree are the number of installed electricity meters per building, its volume and a binary variable indicating the presence of a \ac{PV} installation or a heat pump.
	For the single- and two-family buildings, the number of flats is apparent.
	For the apartment towers, we estimate the exact number of flats based on the number of electricity meters placed in a given tower in a second step.
	We calibrate the results with the latest available census for Germany from 2011 and current data on the number of people living in the considered town.
	
	After the synthesis of the flats, we randomly attribute a family type to every sampled flat according to their relative frequency as they occur in Ha{\ss}furt.
	The available family types are based on the census classes from Germany and include one-person households, couples without children, single parents, couples with children, and multi-person households without nuclear family.
	Once we have assigned a family type, we estimate the number of family members.
	To account for the fact that families with adult children are also classified as families in the German census, we separate the people per household into those with an age of 18 or older and those below 18 based on an estimate of the share of children above 18 years that still live at home.
	Knowing the age of the children is crucial for the following, as in Germany only people aged 18 and over are allowed to drive a vehicle.
	Finally, we have a sampled population of the considered town, including age groups and attributions of the individuals to existing buildings.
	
	In the last step, we attribute the vehicles with combustion engines among the sampled flats/households.
	The total number of vehicles in Ha{\ss}furt is taken from the regionalized statistics on the distribution of vehicles from the \citeN{2022_Zulassungszahlen}, excluding historic or special-purpose cars.
	Based on the results of our survey, we can determine the distribution of the number of vehicles per household based on the number of adults living in that household.
	Using this distribution, we randomly add vehicles to the households according to their number of adults.
	If the sampling results in fewer vehicles per household than there are adults, some adults have to share a vehicle.
	Using this sampling approach, we sample 9~397 vehicles with a combustion engine for~2021.

\subsection{Mobility Demand Generation}
\subsubsection{Mode Choice Model for Vehicle Usage}
\label{sec:methodology_mode_choice}
\begin{figure}[htb]
		{
			\centering
			\includegraphics[width=0.83\textwidth]{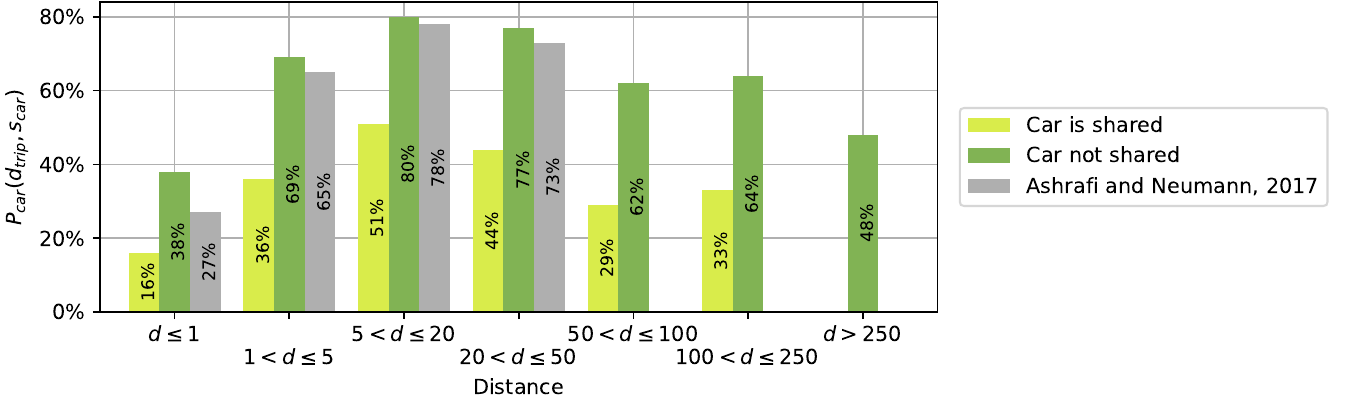}
			\vspace{-7pt}
			\caption{Visualization of the mode choice model that is extracted from the \ac{MiD} dependent on the vehicle sharing type, i.e., vehicle shared between household members (lime) or vehicle exclusively attributed to one person (dark green), including a reference from the literature (gray).\label{fig:method_mode_choice}}
			\vspace{17pt}
		}
	\end{figure}

	The mode choice model indicates which means of transportation is used to complete a trip.
	In this paper, we use a model that returns the probability $P_{car}(d_{trip},s_{car})$ of utilizing a vehicle as a driver based on the distance of the trip $d_{trip}$ and a binary variable $s_{car}$ indicating whether the vehicle is available for the full day for a given driver or if the driver must share the vehicle among other household members.
	We extract the probabilities from the \ac{MiD} survey results, selecting only results from rural counties in Southern Germany as both mode choice and travel distance are strongly influenced by the type of region (i.e., rural or urban) \cite{2010_Scheiner_ModeChoiceAndTripDistancesTrendsGER}.
	In \autoref{fig:method_mode_choice}, we compare our \ac{MiD}-based mode choice model for vehicle usage to related work, analyzing the mode choice for a similar rural region in Austria that is close to our considered area \cite{2017_Ashrafi_ModeChoiceAustria}.
	The comparison is only visualized for the distance groups that merge with our analysis, i.e., up to a trip distance of 50~km.
	
	In this paper, we focus on home-centered tours.
	A home-centered tour is a set of trips where the first trip departs from home and the last trip of the tour arrives at home.
	For the sake of clarity, we assume that the complete tour is executed with the same mode, i.e., intermodal tours are not considered.
	For a given tour, we determine the mode by sampling with the probability $P_{car}(d_{trip},s_{car})$ obtained from the mode choice model using the length of the longest trip of the tour as $d_{trip}$ and the information if another person already uses the vehicle.
	In the literature, we also find other models like logit models based on computing utilities of using different means of transport for a given trip or tour \shortcite{1974_BenAkiva_ModeChoiceModel,2021_Hillel_ModeChoiceModelClassification}.

\subsubsection{Assignment and Processing of Mobility Demand from OMOD to Sampled People}
\label{sec:methodology_OMOD_sampling}
	We generate spatially resolved individual mobility demand profiles using OMOD, the OpenStreetMap Mobility Demand Generator \cite{2023_Strobel_OMOD}.
	OMOD is an open-source tool designed to generate detailed daily activity schedules for a population of agents based on an agent-based simulation calibrated with the \ac{MiD}.
	It provides disaggregated temporal and spatial information at the individual building level, facilitating the easy creation of realistic mobility demand for various applications, including energy systems modeling.
	We apply OMOD to the considered region using a buffer area of 55~km and sample over 10{,}000 adults for one week.
	For every sampled person, OMOD outputs a daily list of trips, including the trip distance, start and arrival time.
	
	The generated individual mobility demand profiles, i.e., the list of trips per person, are postprocessed in multiple steps.
	First, we generate a set of home-centered tours out of the list of trips.
	Therefore, we iterate over the chronologically sorted list of trips starting at home and merge all trips until we reach the home place again.
	To apply the above-described mode choice model, we additionally note the length of the longest trip in every tour.
	Thereupon, we remove all overlapping tours, which happens in some cases as trips of the previous day are not finished on the next day.

\subsection{Vehicle Tour Sampling and Application of the Mode Choice Model}
	The sampling procedure from Section~\ref{sec:methodology_OMOD_sampling} returns home-centered tours per person.
	As our target is to obtain home-centered tours of the vehicles driven by these people, we apply the mode choice model from Section~\ref{sec:methodology_mode_choice} to every tour per person to identify the tours executed by vehicle.
	If a vehicle is shared, the first user is considered to be the primary vehicle user, and his vehicle-based tours are sampled at first using the mode choice model for non-shared vehicles (dark green bars in \autoref{fig:method_mode_choice}).
	For the other users attributed to the vehicle, we use the mode choice model, assuming the vehicle is shared (lime bars in \autoref{fig:method_mode_choice}).
	If this vehicle usage sampling strategy results in overlapping tours, we exclude one of the overlapping tours.
	Also, other papers like \shortciteN{2005_Miller_TourBasedModeChoice} remove one of the overlapping tours to resolve the problem of multiple vehicle users simultaneously, as rescheduling activities lies out of the scope of this paper.

\subsection{Modeling of the \acp{EV} Inside the Digital Twin}
\begin{figure}[htb]
		{
			\centering
			\includegraphics[width=0.7\textwidth]{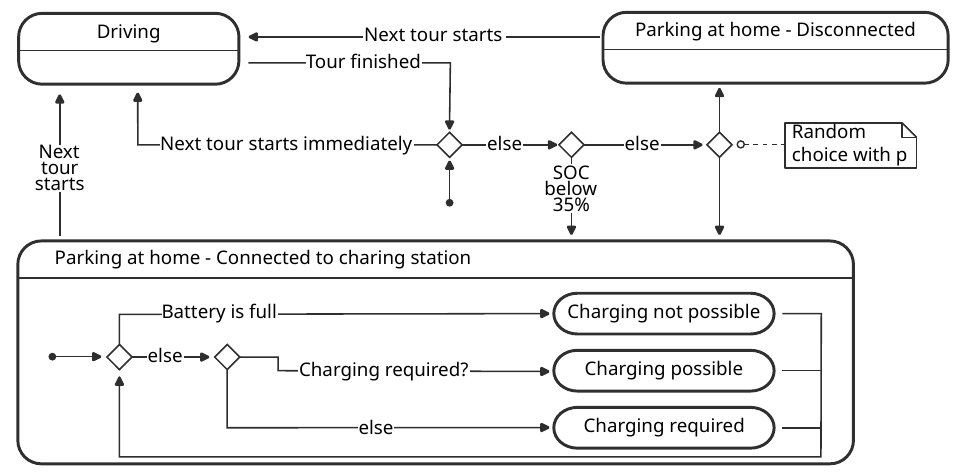}
			\caption{Modeling of the \acp{EV} inside the digital twin as finite-state machine.\label{fig:methodology_state_diagram}}
		}
	\end{figure}

	Within the digital twin, we need the following aspects of \ac{EV} mobility: The charging profile, the connection state to the home charging station, and the question of whether the \ac{EV} is currently parked at home.
	Therefore, we model the \acp{EV} as a finite-state machine having the states that an \ac{EV} is either driving or parked at home.
	In the latter case, it can be connected to the charging station or it may park at home disconnected.
	Changes between the driving and the parking states happen based on the sampled profiles from Section \ref{sec:methodology_OMOD_sampling}.
	If an \ac{EV} is parked once, the connection state cannot change anymore.
	The connection state is determined directly after the \ac{EV} has arrived based on the random choice model obtained from the survey (see Section \ref{sec:methodology_survey_charging_choice}).
	If the battery state is below 35\%, the \ac{EV} is always connected to the charging station.
	If the \ac{EV} is parked and connected, the following sub-states describing the charging possibilities are possible: Either the battery is fully charged, so no additional charging is possible.
	If the battery can still be charged, we distinguish two cases.
	Either the \ac{EV} must be charged so that the next tour can be finished without additional charging during the tour, or charging is possible but not mandatory.
	The complete state diagram, including all possible transitions as described above, is depicted in \autoref{fig:methodology_state_diagram}.
	Additionally, we assume that the vehicles require 
	20~kWh/100~km, which is the upper value reported by \citeN{2019_Simon_ModellingIncreaseOfEV}.
	
\subsection{Possible Addition of Rooftop \ac{PV} Installations and \ac{BESS}}
\label{sec:method_PV_addition}
	To analyze the self-consumption potential of rooftop \ac{PV} generation for charging an \ac{EV}, we size the \ac{PV} installations according to the available roof area, as this is an inherent feature of the applied digital twin.
	We assume that it is possible to install $0.172$~kWp/m\textsuperscript{2} of roof area, with an upper limit of 30~kWp as defined by the German tax legislation.
	The used profiles are extracted from already existing, individually metered \ac{PV} installations grouped according to their orientation in the area under consideration.
	
	For the simulation of added residential \acp{BESS}, we use a power-energy model \shortcite{2022_Vykhodtsev_ModellingApproachesForLiIonBESS} with a round-trip efficiency of 90\% and no self-discharge.
	The simulated battery's kWh capacity matches the building's annual electricity usage in MWh (excluding \acp{EV}), capped at a maximum of 20~kWh.
	A rule-based control strategy governs the charging or discharging of the added \ac{BESS} to optimize \ac{PV} self-consumption (see \shortciteN{2023_Bayer_DT_LocalEnergySystem} for details).

\subsection{Internal Logic of the Digital Twin and Analyzed Metrics}
\label{sec:method_internal_logic_and_metrics}
	For the analysis of the share of \ac{EV} charging demand that can be covered by a residential rooftop \ac{PV} installation, we analyze the metrics described below.
	For this analysis, we assume that the residential electricity demand has priority for the \ac{PV} self-consumption and that there is no additional charging other than at the home of the \ac{EV}.
	The main focus is on the analysis of the \ac{SCR} and the \ac{SSR} that are defined according to \shortciteN{2016_PV_BESS_SCR_SSR_Sweden} for a building $c$ by
	\begin{equation}
		SCR = \frac{\text{Sum of self-consumed PV energy}}{\text{Sum of total PV generation}} = \frac{\sum_{t\in T}P^{\text{self cons.}}_c(t)}{\sum_{t\in T}P^{\text{PV}}_c(t)}
	\end{equation}
	\begin{equation}
		SSR = \frac{\text{Sum of self-consumed PV energy}}{\text{Sum of total demand}} = \frac{\sum_{t\in T}P^{\text{self cons.}}_c(t)}{\sum_{t\in T}P^{\text{build}}_c(t) + P^{\text{CS}}_c(t)}
	\end{equation}
	where $P^{\text{build}}_c(t)$ denotes the residential electricity demand of the building at time step $t$, $P^{\text{PV}}_c(t)$ is the \ac{PV} generation at $t$, $P^{\text{CS}}_c(t)$ is the demand of the charging station at $t$ and $P^{\text{self cons.}}_c(t)$ denotes the self-consumed power at time step $t$.
	For the computation of the self-consumed power $P^{\text{self cons.}}_c(t)$ at a given time step $t$ and a building $c$, we extend the computation as presented in \citeN{2023_Bayer_DT_LocalEnergySystem} by the power of the residential \ac{EV} charging station:
	\begin{align}\label{eq:P_self_cons}
		P^{\text{self cons.}}_c(t) =
		\min \{ \; \underbrace{P^{\text{build}}_c(t) + P^{\text{CS}}_c(t)}_{\text{Local demand}},\;\; \underbrace{P^{\text{PV}}_c(t) - P^{\text{Bat}}_{act,c}(t)}_{\text{Local production}} \; \}
	\end{align}
	Please note that $P^{\text{PV}}_c(t) - P^{\text{Bat}}_{act,c}(t) \geq 0$ holds always following as argued in \citeN{2023_Bayer_DT_LocalEnergySystem}.

\section{RESULTS}
\label{sec:results}
	In this section, we first present the aggregated results of the generation of vehicle usage profiles.
	To be able to use these profiles for the computation of \ac{EV} charging profiles assuming immediate charging when the \ac{EV} arrives at home and no charging on the way, we need to verify that the sampled vehicle profiles correctly represent the vehicle usage behavior.
	Therefore, we compare our results to other sources or questions of our survey that have not been used in the methodology.

\subsection{Results and Validation of the Vehicle Usage Sampling}
\label{sec:results_validation_of_sampling}
\begin{figure}[htb]
		{
			\centering
			\includegraphics[width=0.98\textwidth]{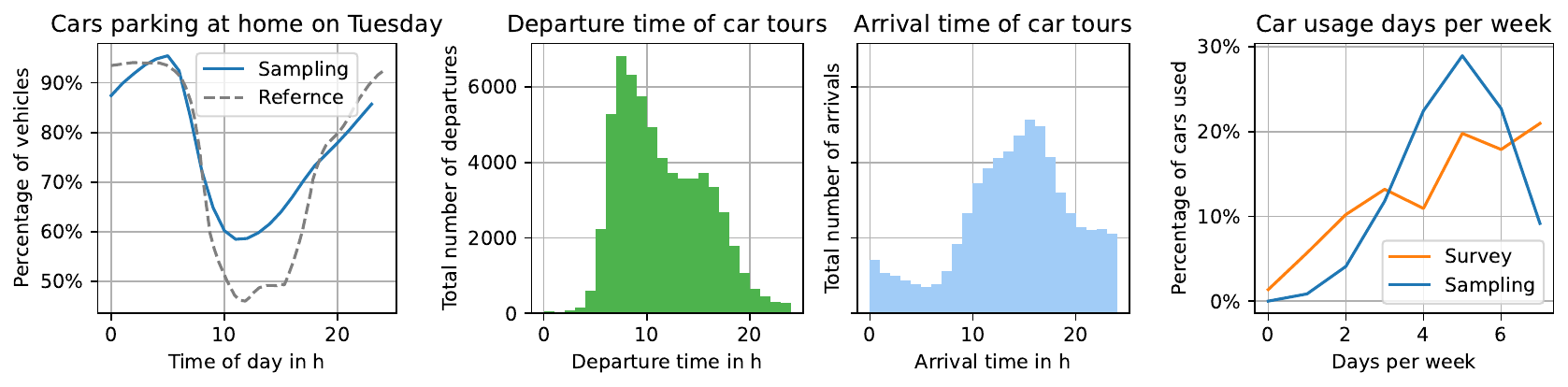}
			\caption{Aggregated output of the mobility demand processing. The blue line in the leftmost plot shows the percentage of vehicles parking at home for an exemplary weekday (Tuesday) as simulated. The two plots in the center visualize the histogram of the departure and arrival times of all vehicle tours over the complete sampled week. The rightmost plot shows the number of days a vehicle is used per week.\label{fig:results_car_usage}}
		}
	\end{figure}

	To justify that our sampled vehicle usage profiles represent the usage patterns of vehicles within a weekday, we first compare the share of vehicles parking at home in our sample (blue line in  \autoref{fig:results_car_usage}) with a travel survey from the United Kingdom, as analyzed by \shortciteN{2018_Crozier_MobilitySurveysForEVProfileGen} (dashed gray line).
	The comparison shows that our sample and the reference exhibit the same trend over the Tuesday as an exemplary day of the week.
	Nevertheless, as we consider another region as the reference, a direct comparison of the percentage of vehicles parked at home is impossible.
	In \autoref{fig:results_car_usage}, we also include the daily departure and arrival times of vehicles at their home place.
	Even though we cannot find a survey that evaluates these times in a European country, these histograms seem to be plausible.
	In the morning, most vehicles depart between 6~a.m. and 9~a.m., and in the evening, most vehicles arrive between 15~a.m. and 16~a.m., which reflects a typical workday usage.
	This leads to the conclusion that our methodology is capable of correctly simulating vehicle usage within a day.

	To verify that the weekly vehicle usage is correctly represented, we compare the number of days a week the vehicle is used at least one time that day between our sampled profiles (blue line in the rightmost plot from \autoref{fig:results_car_usage}) and the survey results from our conducted survey (orange line).
	Most vehicles in our simulation are used five times a week (29\%).
	However, the survey suggests that most vehicles are used daily (21\%).
	On the other hand, our simulation shows that there are hardly any vehicles with less than two days of use per week (less than 5\%), which was indicated in the survey in over 17\% of cases.
	Nevertheless, the average number of vehicle usage days per week for all vehicles is 4.6 in the survey and 4.8 in our sample.
	This leads to the conclusion that our sampled vehicle usage profiles represent the average vehicle usage, but do not perfectly cover the complete range of diverse vehicle usage.

\subsection{Analysis of \ac{PV} Self-Consumption for Residential \ac{EV} Charging}
	To analyze the \ac{PV} self-consumption for residential \ac{EV} charging, we consider all residential buildings in Ha{\ss}furt that   have neither a \ac{PV} installation nor an \ac{EV}.
	In total, these are 3~353 residential buildings.
	We define the following scenarios that are evaluated using \ac{PV} generation and smart meter data from~2021 in our existing digital twin \cite{2023_Bayer_DT_LocalEnergySystem} with an hourly resolution:
	\begin{itemize}
	\setlength\itemsep{0pt}
		\item \textbf{CS:} Baseline scenario, i.e., the current state without additional components like \ac{EV} charging stations of \ac{PV} installations
		\item \textbf{EV:} All residential buildings are equipped with an \ac{EV} charging station
		\item \textbf{PV:} All residential buildings are equipped with a \ac{PV} installation
		\item \textbf{PV+BS:} All residential buildings are equipped with a \ac{PV} installation and a \textit{BESS}
		\item \textbf{EV+PV:} Combination of scenario \textit{EV} and \textit{PV}
		\item \textbf{EV+PV+BS:} Combination of scenario \textit{EV}, \textit{PV+BS}
	\end{itemize}
	The main reason for including the scenarios \textit{PV} and \textit{PV+BS} is to show the additional effect of the \acp{EV}.
	All added \ac{PV} installations and \acp{BESS} are sized as stated in Section~\ref{sec:method_PV_addition}.
	
\begin{figure}[htb]
		{
			\centering
			\includegraphics[width=\textwidth]{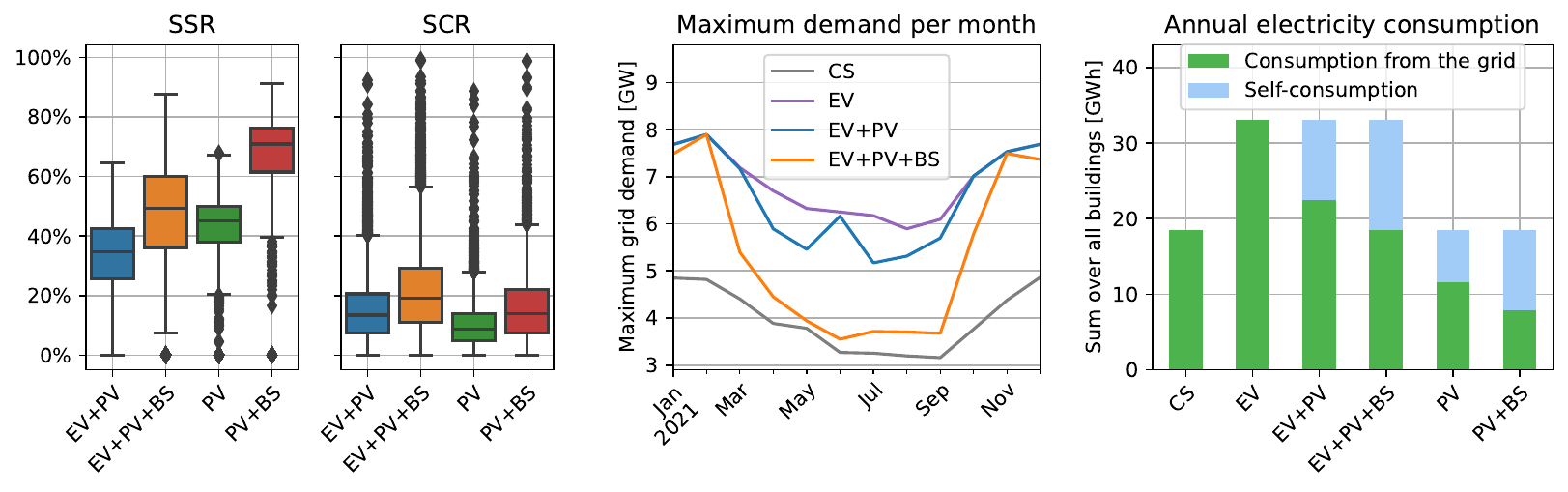}
			\caption{Distribution of SSR and SCR (left plots) over all considered buildings with added \acp{EV} in the scenarios with EV-addition. Center plot: Maximum residential demand per month for different scenarios. Rightmost plot: Sum of annual electricity consumption over all considered buildings, separated by the actual electricity consumed from the local grid (green) and the consumption covered by the local rooftop PV installation (light blue).\label{fig:results_SSR_SCR_SumE}}
		}
	\end{figure}

	The results show that self-sufficiency is massively reduced in the presence of \acp{EV} compared to a scenario without \acp{EV} (see leftmost plot in \autoref{fig:results_SSR_SCR_SumE}).
	The average \ac{SSR} reduces from 46\% in Scenario \textit{PV} to 38\% in Scenario \textit{EV+PV}.
	On the other hand, the annually self-consumed electricity increases in the presence of \acp{EV}.
	The average \ac{SCR} increases from 13\% in Scenario \textit{PV} to 18\% in Scenario \textit{EV+PV}.
	Similar changes in the values of \ac{SSR} and \ac{SCR} occur in the presence of \acp{BESS}.
	
	Summed over all buildings, adding \acp{EV} to all considered buildings leads to an increase in the annual electricity consumption from 18.5~GWh to 33.0~GWh, corresponding to an increase of 78\% (see rightmost plot in \autoref{fig:results_SSR_SCR_SumE}).
	With the addition of a rooftop \ac{PV} installation to all buildings, we can reduce the annual electricity consumed from the grid by 31.2\% to 22.5~GWh.
	When further adding a \ac{BESS} (Scenario \textit{EV+PV+BS}), the summed annual grid consumption reduces to a value of 18.4~GWh, which is even 0.1~GWh less than the current state (CS).
	This means that we can eliminate the added annual consumption of the \acp{EV} by adding rooftop \ac{PV} installations and \acp{BESS} to all buildings.
	Nevertheless, the comparison to a scenario where no \acp{EV} are integrated reveals that the charging of the simulated \acp{EV} still requires grid demand as the total consumption from the grid of all buildings without \acp{EV} is much lower (11.6~GWh in Scenario \textit{PV} and 7.9~GWh in Scenario \textit{PV+BS}).
	The monthly peak demand accumulated over all buildings is highly affected by adding \acp{EV} (see center plot in \autoref{fig:results_SSR_SCR_SumE}).
	The peak increases from 4.9~GW in the current state to 7.9~GW.
	The addition of \ac{PV} installations can reduce this peak during most of the summer months by around 1~GW.
	However, most of the peaks can be reduced using additional \acp{BESS}, especially from April until September.

\section{DISCUSSION AND LIMITATIONS}
\label{sec:discussion_and_limitations}

	The validation of the vehicle usage profiles, as shown in Section~\ref{sec:results_validation_of_sampling}, shows that our presented methodology is capable of generating realistic vehicle driving profiles.
	Regarding the results for the second research question, we can highlight that combining an \ac{EV} with a \ac{PV} installation and \ac{BESS} for every considered building yields a substantial reduction in the annual building electricity consumption, nearly matching the current state without additional \acp{EV}, \ac{PV} installations, or \ac{BESS}.
	This integrated approach demonstrates significant potential for enhancing energy efficiency and sustainability within the transportation sector while leveraging renewable energy sources to offset traditional grid reliance.
	Compared to existing papers analyzing the \ac{SCR} of rooftop \ac{PV} installations in Germany, which resulted in values between 26\% and 38\% (without storage) \cite{2022_Hassan_SelfConsumptionOfPVSystems}, most of the analyzed buildings in our situation show lower \ac{SCR} values.
	This is due to the fact that the PV systems considered in our case sized according to the available roof area are considerably larger (21~kWp in the mean) than in the considered paper (between 1.1 and 6~kWp).
	
	A methodological limitation of our paper is the sampling process.
	Existing literature, such as \shortciteN{2018_Eisenmann_DifferentCarUsageGermanyVSCalifornia}, reports that people usually use the same means of transportation for repeating activities, like going to work or the same shopping location.
	In the current implementation, we sample the means of transport for every occurrence of a possibly repeating tour again.
	This suggests that the simulation incorrectly estimates the use of the vehicle for repeated trips over a week.
	Nevertheless, the estimation of the average vehicle usage days per week represents the vehicle usage correctly (see Section \ref{sec:results_validation_of_sampling}).
	Moreover, our results are limited to the area considered, including the local weather conditions, in this case, Southern Germany.

\section{CONCLUSION}
\label{sec:conclusion}
	This paper presents a novel combination of a local energy system's digital twin and a mobility simulation to evaluate the impact of higher \ac{EV} penetration rates in future energy system states.
	To integrate the mobility simulation, we incorporated results from an additional survey.
	We verify the presented methodology for generating vehicle mobility profiles from sampled human mobility demands by comparing the weekly and intra-day vehicle usage patterns to other publications or additional survey results.
	As an example of application-level results, we evaluate the possibility of compensating for the additional electricity consumption caused by EVs in the private sector.

	Our results offer multiple insights at various levels.
	On a methodological level, we prove that it is possible to combine a real-world digital twin with a mobility simulation on an individual level if additional information about the structure of the households (number of adults/children per household), the households/flats per building and the vehicles per household are present.
	As these data are not fully covered by existing surveys or data from the statistical offices, we added survey information from a survey we conducted.
	At the application level, the results show that adding \acp{EV} increases the total annual electricity consumption across all considered residential buildings by 78\% with home charging only.
	By adding PV systems and \acp{BESS} to all buildings under consideration, the annual electricity consumption of these buildings can be reduced back to the current level without \acp{EV}, \ac{PV} systems or \acp{BESS}.
	Similarly, the demand peaks caused by \ac{EV} charging can be almost entirely prevented by the addition of \ac{PV} systems and \acp{BESS} in the summer months.
	All in all, the negative effects of increasing \ac{EV} penetration on the energy system can only really be mitigated by additional \acp{BESS}.
	As both systems, \acp{BESS} and smart charging infrastructure can be costly, it is promising to compare the Scenario \textit{EV+PV} using additional smart charging with Scenario \textit{EV+PV+BS} from a financial point of view for future research.

\section*{ACKNOWLEDGMENTS}
We would like to thank our research partner Stadtwerk Hassfurt GmbH for providing the data set.
This paper is an outcome of the research project \textit{DigiSWM} (DIK-2103-0017 / DIK0298/02) founded by the Bavarian State Ministry of Economic Affairs, Regional Development and Energy.

\footnotesize

\bibliographystyle{wsc}
\bibliography{referencesP05}

\section*{AUTHOR BIOGRAPHIES}

\noindent {\bf \MakeUppercase{Daniel Bayer}} received his Master's degree in Computer Science in 2021. 
Currently, he is a PhD Student 
interested in digital twins and data-driven simulations of local energy systems, including the building sector, to ensure long-term sustainable and climate-neutral energy supply through the optimal dimensioning and control of heating and energy supply systems.
His email address is \email{daniel.bayer@uni-wuerzburg.de} and his ORCID-ID is 0000-0002-4063-4097.\\

\noindent {\bf \MakeUppercase{Marco Pruckner}} holds the Chair of Modeling and Simulation at the Institute of Computer Science at the University of W\"urzburg. His research interests include modeling and simulation of energy and mobility systems, simulation optimization, machine learning and data analytics. His e-mail address is \email{marco.pruckner@uni-wuerzburg.de}.\\

\end{document}